\documentclass[useAMS,usenatbib]{mn2e}

 \usepackage{graphicx}

\title[Does a particle accelerator arise near the light cylinder?]
{Does a strong particle accelerator arise very close to 
the light cylinder in a pulsar magnetosphere?}
\author[K. Hirotani]{K. Hirotani$^{1}$\thanks{E-mail:
hirotani@tiara.sinica.edu.tw}\\ 
$^{1}$ Academia Sinica, Institute of Astronomy and Astrophysics (ASIAA),
       PO Box 23-141, Taipei, Taiwan}
\begin{document}

\date{Accepted ... Received 2014 January 10; 
in original form 2014 January 10}

\pagerange{\pageref{firstpage}--\pageref{lastpage}} \pubyear{2002}

\maketitle

\label{firstpage}

\begin{abstract}
We examine if an efficient particle acceleration
takes place by a magnetic-field-aligned electric field 
near the light cylinder in a rotating neutron star magnetosphere.
Constructing the electric current density with the
actual motion of collision-less plasmas,
we express the rotationally induced, Goldreich-Julian charge
density as a function of position.
It is demonstrated that the \lq light cylinder gap',
which emits very high energy photons via curvature process
by virtue of a strong magnetic-field-aligned 
electric field very close to the light cylinder,
will not arise in an actual pulsar magnetosphere.
\end{abstract}

\begin{keywords}
   acceleration of particles
-- gamma rays: stars
-- magnetic fields
-- methods: analytical
-- stars: neutron
\end{keywords}

\section{Introduction}
The Crab pulsar (PSR~J0534+2200),
one of the youngest pulsars in our Galaxy,
shows pulsed signals in a very wide energy range
from radio to $\gamma$-rays
(e.g., see Abdo et al. 2010 for the observation
 of this pulsar with {\it Fermi} LAT between 100~MeV and 20~GeV).
In the highest energy range,
the Major Atmospheric Gamma-Ray Imaging Cherenkev (MAGIC) telescope
has detected pulsed signals at 25~GeV (Aliu et al. 2008),
which was confirmed by the LAT observations
(Atwood et al. 2009).
Further observations with the MAGIC telescope and 
the Very Energetic Radiation Imaging Telescope Array System (VERITAS)
have shown that this component extends up to 400~GeV
(Aleksi\'{c} et al. 2012; Aliu et al. 2011).

To explain such pulsed fluxes in the very high energy (VHE) region
(i.e., above 100~GeV),
Bednarek (2012) proposed the \lq light cylinder (LC) gap' model
from the following reasons:
The rotationally induced, Goldreich-Julian (GJ) charge density
is given by (Goldreich-Julian~1969)
$ \rho_{\rm GJ}
  = -\left[ \mbox{\boldmath$\Omega$}\cdot\mbox{\boldmath$B$}/(2\pi c)
     \right]
     \left[ 1-(\varpi/\varpi_{\rm LC})^2 
     \right]^{-1},
$
where $\mbox{\boldmath$\Omega$}$ denotes the rotation vector of 
the neutron star, 
$\Omega=\vert\mbox{\boldmath$\Omega$}\vert$ its rotation frequency,
$\mbox{\boldmath$B$}$ the magnetic field at each point,
$\varpi$ the distance from the rotation axis, 
$\varpi_{\rm LC}=c/\Omega$ the radius of the LC
measured from the rotation axis, and $c$ the speed of light.
If the real charge density, $\rho_{\rm r}$, coincides $\rho_{\rm GJ}$ 
at every position,
the magnetic-field-aligned electric field, $E_\parallel$, 
vanishes in the entire region of the magnetosphere. 
If $\rho_{\rm r}$ deviates from $\rho_{\rm GJ}$ at some position,
on the other hand,
the acceleration electric field $E_\parallel$ will arise
around that position.
In this expression, $\rho_{\rm GJ}$ appears to diverge at the LC, 
$\varpi \rightarrow \varpi_{\rm LC}$; thus,
it was argued if $\rho_{\rm r}$ inevitably deviates from
$\rho_{\rm GJ}$ near the LC.
By virtue of this diverging behavior of $\rho_{\rm GJ}$,
an extremely strong $E_\parallel$,
which is about $10^3$ stronger than what arises in the 
outer-magnetospheric particle accelerator (or the outer gap),
was assumed to arise in the LC gap, 
and the resultant curvature emission was implied to reproduce
the pulsed spectrum observed from the Crab pulsar up to 400~GeV.

In \S~\ref{sec:GJ}, we demonstrate that the LC gap model is not feasible,
examining the actual $\rho_{\rm GJ}$ distribution.
Then in \S~\ref{sec:disc}, we briefly mention an appropriate way
to compute $\rho_{\rm GJ}$.

\section[]{Goldreich Julian charge density in the outer magnetosphere}
\label{sec:GJ}

In the special relativistic limit, the GJ charge density
is given by (e.g., Mestel \& Wang~1982)
\begin{equation}
  \rho_{\rm GJ} 
  \equiv -\frac{\mbox{\boldmath$\Omega$}\cdot\mbox{\boldmath$B$}}
               {2\pi c}
         +\frac{(\mbox{\boldmath$\Omega$}\times\mbox{\boldmath$r$})\cdot
                (\nabla\times\mbox{\boldmath$B$})}
               {4\pi c}.
  \label{eq:def_rhoGJ_1}
\end{equation}
From the inhomogeneous part of the Maxwell equations,
we obtain,
\begin{equation}
  \nabla\times\mbox{\boldmath$B$}
  = \frac{4\pi}{c} \mbox{\boldmath$J$}
   +\frac{1}{c}\frac{\partial\mbox{\boldmath$E$}}{\partial t}.
  \label{eq:divB}
\end{equation}
Since the plasmas are highly collision-less in a pulsar magnetosphere,
charged particles gyrate many times between collisions. 
Thus, we must construct the electric current $\mbox{\boldmath$J$}$ 
from the gyrating and drifting motion of charged particles, 
not from the generalized Ohm's law.
Let us decompose the current into 
the parallel and perpendicular components with respect to 
the local magnetic field line, 
\begin{equation}
  \mbox{\boldmath$J$}
  = \mbox{\boldmath$J$}_\parallel+\mbox{\boldmath$J$}_\perp.
  \label{eq:J1}
\end{equation}

First, we consider the parallel current.
Since the radiation force balances with the electrostatic acceleration,
particles' distribution becomes mono-energetic.
Thus, denoting the terminal velocity of 
out-going particles (e.g., positrons) 
with $\mbox{\boldmath$v$}_{\parallel +}$, and
in-going ones (e.g., electrons) 
with $\mbox{\boldmath$v$}_{\parallel -}$, 
we obtain
\begin{equation}
  \mbox{\boldmath$J$}_\parallel
  = e( n_+ \mbox{\boldmath$v$}_{\parallel +} 
      -n_- \mbox{\boldmath$v$}_{\parallel -})
  \label{eq:J2}
\end{equation}
where $n_+$ (or $n_-$) denotes the number density 
of out-going (or in-going) particles,
and $\mbox{\boldmath$v$}_{\parallel \pm}$ is given by
(Hirotani 2011, ApJ 733, L49)
\begin{eqnarray}
  \mbox{\boldmath$v$}_{\parallel \pm}
  &=& c f_\pm \frac{\mbox{\boldmath$B$}}{B},
  \nonumber\\
  f_\pm
  &\equiv& 
      -\frac{\varpi}{\varpi_{\rm LC}}\frac{B^{\hat\varphi}}{B}
         \pm \sqrt{1-\left(\frac{\varpi}{\varpi_{\rm LC}}\right)^2
                     \left(\frac{B_{\rm p}}{B}\right)^2};
  \label{eq:J3}
\end{eqnarray}
$e$ denotes the charge on the out-going particle 
(presumably the positron),
$B_{\rm p}^2=B^2-(B^{\hat\varphi}{})^2$,
and $B^{\hat\varphi}$ the toroidal component of the magnetic field.
In the higher altitudes (e.g., near the LC),
it is reasonable to assume $n_+ \gg n_-$ in the gap.
In this case, the real charge density is given by
$\rho_{\rm r}=e(n_+-n_-) \approx e n_+$.
Thus, we obtain
\begin{equation}
  \mbox{\boldmath$J$}_\parallel
  = \rho_{\rm r} \mbox{\boldmath$v$}_{\parallel +}
  \label{eq:J4}
\end{equation}
Even if $n_- \approx n_+$, the additional term that would appear
in the right-hand side will not change the entire discussion
of this paper;
however, we assume $n_- \ll n_+$ to clarify the logic.

Second, we consider the perpendicular current.
It is given by 
\begin{eqnarray}
  \mbox{\boldmath$J$}_\perp
  &=& c \rho_{\rm r} 
     \frac{\mbox{\boldmath$E$}\times\mbox{\boldmath$B$}}{B^2}
    + c(P_\perp+P_\parallel) 
      \frac{\mbox{\boldmath$B$}}{B^2} \times \frac{\nabla B}{B} 
 \nonumber\\
  && + \frac{c^2 \rho}{B^2}\dot{\mbox{\boldmath$E$}}_\perp
     -c\nabla \times \left( P_\perp \frac{\mbox{\boldmath$B$}}{B^2} 
                     \right),
  \label{eq:J5}
\end{eqnarray}
where $P_\parallel$ and $P_\perp$ denote
the pressure associated with the longitudinal and perpendicular motion
with respect to the magnetic field;
$\rho$ (in the second line) denotes the mass density of the plasmas,
and $\dot{\mbox{\boldmath$E$}}_\perp$
the temporal derivative of the electric field projected on the
perpendicular plane to $\mbox{\boldmath$B$}$.
In the right-hand side, 
the first term represents the current due to the $E \times B$ drift,
the second term the sum of the currents due to
the magnetic-gradient and the magnetic-curvature drift,
the third term (in the second line) the polarization-drift current,
and the last term the magnetization current.
In a collision-less plasma, the pressure tensor becomes 
highly anisotropic.
In a pulsar magnetosphere, pairs are created inwards
(via photon-photon and/or magnetic pair creation
 in the middle or lower altitudes) 
with the typical Lorentz factor of a few thousand.
Thus, positrons (or electrons) lose most of their perpendicular 
momentum when they return outwards by a positive (or a negative)
$E_\parallel$ in a strong magnetic field.
Moreover, their pitch angles decrease due to a subsequent acceleration
by $E_\parallel$,
resulting in $P_\perp \ll P_\parallel$.
Thus, for particles migrating in the outer magnetosphere,
we obtain 
\begin{equation}
  \mbox{\boldmath$J$}_\perp
  = c \rho_{\rm r} \frac{\mbox{\boldmath$E$} \times \mbox{\boldmath$B$}}
                       {B^2}
   +c P_\parallel \frac{\mbox{\boldmath$B$}}{B^2}
                \times \frac{\nabla B}{B}
   + \frac{c^2 \rho}{B^2}\dot{\mbox{\boldmath$E$}}_\perp.
  \label{eq:J7}
\end{equation}

In a co-rotating magnetosphere, 
we can put $\rho_{\rm r}=\rho_{\rm GJ}$
and have
$c \mbox{\boldmath$E$} \times \mbox{\boldmath$B$}
 = (\mbox{\boldmath$\Omega$} \times \mbox{\boldmath$r$}) B^2$.
Thus, combining equations~(\ref{eq:def_rhoGJ_1}), 
(\ref{eq:divB}), (\ref{eq:J1}), (\ref{eq:J4}), and (\ref{eq:J7}), 
we obtain
\begin{eqnarray}
  \lefteqn{
  \left[ 1-\left( f_+ \frac{B^{\hat\phi}}{B}
                 +\frac{\gamma m_{\rm e} c^2}{eB}
                  \frac{b^{\hat\varphi}}{L}
                 +\frac{m_{\rm e}c \dot{E}^{\hat\varphi}}
                       {eB^2}
           \right)
           \frac{\varpi}{\varpi_{\rm LC}}
          -\left(\frac{\varpi}{\varpi_{\rm LC}}\right)^2 
  \right]
  \rho_{\rm GJ}
  }
  \nonumber\\
  &=& -\frac{\mbox{\boldmath$\Omega$}\cdot\mbox{\boldmath$B$}}
            {2\pi c}
      +\frac{\mbox{\boldmath$e$}_{\hat\varphi}\cdot\dot{E}}
            {4\pi c}
       \frac{\varpi}{\varpi_{\rm LC}},
  \label{eq:rho_1}
\end{eqnarray}
where
\begin{equation}
  \frac{b^{\hat\varphi}}{L}
  \equiv          \mbox{\boldmath$e$}_{\hat\varphi}
                  \cdot \left( \frac{\mbox{\boldmath$B$}}{B}
                               \times \frac{\nabla B}{B}
                        \right),
\end{equation}
$L \sim \varpi_{\rm LC}$, $\vert b^{\hat\varphi} \vert \sim 1$,
and $\mbox{\boldmath$e$}_{\hat\varphi}$ denotes the toroidal 
unit vector.
Note that $\gamma m_{\rm e} c^2$ is much small compared to $eBL$
if particles efficiently radiate,
and  that $\vert \dot{E}^{\hat\varphi} \vert < \Omega B$.
We thus finally obtain
\begin{equation}
  \rho_{\rm GJ}
  = \frac{\displaystyle{-\frac{\mbox{\boldmath$\Omega$}\cdot
                               \mbox{\boldmath$B$}}
                              {2\pi c}
                        +\frac{\mbox{\boldmath$e$}_{\hat\varphi}
                               \cdot\dot{E}}
                              {4\pi c}
                         \frac{\varpi}{\varpi_{\rm LC}}
                       }
         }
         {\displaystyle{ 1-f_+ \frac{B^{\hat\phi}}{B}
                               \frac{\varpi}{\varpi_{\rm LC}} 
                        -\left(\frac{\varpi}
                                    {\varpi_{\rm LC}}
                         \right)^2 
                       }
         }
  \label{eq:rho_2}
\end{equation}
In the numerator, 
the second term is usually small compared to the first term.
In the denominator,
we should notice that $f_+$ is positive definite, 
provided $B^{\hat\phi}<0$.  
For example, at the LC, we obtain 
$f_+=2\vert B^{\hat\varphi}\vert/B$.
Thus, we find
\begin{equation}
  - f_+ \frac{B^{\hat\phi}}{B}
        \frac{\varpi}{\varpi_{\rm LC}}
  > 0.
  \label{eq:rho_3}
\end{equation}
It follows that the denominator of equation~(\ref{eq:rho_2})
does not vanish at the LC, 
provided that the magnetic field is toroidally bent.
Moreover, near the LC, we obtain
\begin{equation}
  - f_+ \frac{B^{\hat\phi}}{B}
        \frac{\varpi}{\varpi_{\rm LC}}
  \approx 1.
\end{equation}
Therefore, the GJ charge density 
is kept around its Newtonian value, 
$-\mbox{\boldmath$\Omega$}\cdot\mbox{\boldmath$B$}/(2\pi c)$,
even near the LC.

\begin{figure}
 \includegraphics[width=8.0cm]{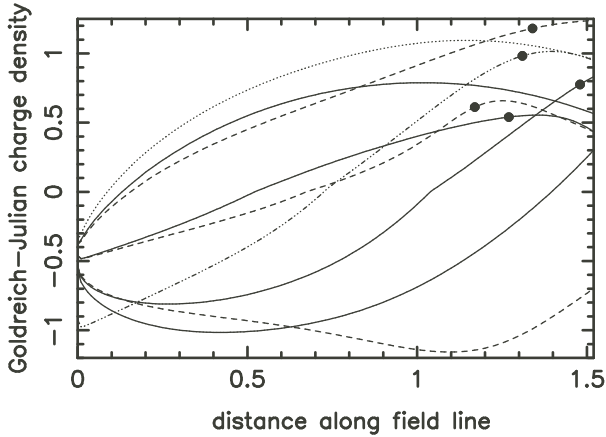}

 \caption{
Distribution of the dimensionless Goldreich-Julian charge density,
$\rho_{\rm GJ}/(\Omega B/2\pi c)$,
as a function of the distance along the 
last-open magnetic field line,
for discrete values of azimuthal angles, $\varphi_\ast$, 
measured counter-clockwise around the magnetic axis
on the polar-cap surface.
Magnetic inclination angle is assumed to be $60^\circ$
{
for the solid, dashed, and dotted curves, while
$0^\circ$ for the dot-dot-dot-dashed one.
}
The solid curves represent $\rho_{\rm GJ}/(\Omega B/2\pi c)$
in the trailing side of the rotating magnetosphere
{
(from the top,  $\varphi_\ast= -45^\circ$, $\varphi_\ast= -90^\circ$,
$\varphi_\ast= -135^\circ$, and $\varphi_\ast= -180^\circ$)
}, while
the dashed ones in the leading side
{
(from the top,  $\varphi_\ast= -45^\circ$, $\varphi_\ast= -90^\circ$,
and $\varphi_\ast= -135^\circ$)
}.
The rotational and magnetic axes, as well as the
footpoints of the magnetic field lines of 
$\varphi_\ast=180^\circ$ and $\varphi_\ast=0^\circ$
at the polar-cap surface,
reside on the same meridional plane.
From the magnetic pole, the direction $\varphi_\ast=180^\circ$
points the rotation axis, while $\varphi_\ast=0^\circ$ the equator.
The filled circle denotes the position at which the distance
from the rotation axis becomes the light cylinder radius.
}
 \label{fig:rhoGJ}
\end{figure}

We can confirm this result by substituting the solution of 
the vacuum, rotating dipole magnetic field (Cheng et al. 2000) 
into equation~(\ref{eq:def_rhoGJ_1}).
In figure~\ref{fig:rhoGJ},
we plot $\rho_{\rm GJ}/[\Omega B/(2\pi c)]$ as a function of
the distance along each magnetic field line.
{
The magnetic inclination angle $\alpha$ between
the magnetic and rotational axes, is assumed to be $60^\circ$ 
for the solid, dashed, dotted curves, whereas
$0^\circ$ for the dot-dot-dot-dashed one.
}
The solid (or dashed) curves show the results in 
the trailing (or leading) side of the rotating magnetosphere.
The filled circle denotes the position at which 
the field line crosses the light cylinder.
It is confirmed by this explicit calculation
that $\rho_{\rm GJ}$ is kept around its 
Newtonian value even near the LC.

{
The conclusion is unchanged for small inclination angles.
Adopting the vacuum rotating magnetic dipole solution, 
we obtain $B^{\hat\phi}=0$ if $\alpha=0^\circ$.
As a result, equation~(\ref{eq:rho_2}) appears to give 
a diverging $\rho_{\rm GJ}$ at the LC.
Nevertheless, the vacuum solution gives 
$\nabla\times\mbox{\boldmath$B$}=0$ when $\alpha=0^\circ$.
Thus, equation~(\ref{eq:def_rhoGJ_1}) shows
that $\rho_{\rm GJ}$ exhibits no singular behavior at the LC.
We plot the case of $\alpha=0^\circ$ 
as the dot-dot-dot-dashed curve in figure~\ref{fig:rhoGJ},
calculating equation~(\ref{eq:def_rhoGJ_1}) 
from the vacuum, rotating dipole solution.
It follows that $\rho_{\rm GJ}$ does not change rapidly at the LC
also for an aligned rotator, as expected.
}

{
Recently,
}
Bednarek (2012) assumed that $\rho_{\rm GJ}$ becomes as large as
$\sim 10^3 \Omega B/(2\pi c)$
in a short length
$\sim 10^{-3} \varpi_{\rm LC}$ along the magnetic field line
in the vicinity of the LC,
and considered the curvature radiation that reproduces
the pulsed emissions up to 400~GeV from the Crab pulsar.
However, since $\rho_{\rm GJ}$ is kept of the order of
$\Omega B/(2\pi c)$ even at the LC,
this assumption cannot be justified.
In another word, the light cylinder gap,
which is suggested to produce the pulsed VHE emission from the Crab pulsar, 
does not appear in any pulsar magnetosphere.

\section{Discussion}
\label{sec:disc}
%
{
We arrive at the conclusion that the Goldreich-Julian charge
density does not show any singular behavior at the light cylinder.
We may note, in passing, that 
}
the Goldreich-Julian charge density 
should be computed from equation~(\ref{eq:def_rhoGJ_1}) directly,
using the given magnetic field distribution
in the three-dimensional rotating magnetosphere,
instead of replacing  
$\nabla\times\mbox{\boldmath$B$}$ with the current
(i.e., instead of using eq.~[\ref{eq:rho_2}]).
We should notice here that we derive 
equation~(\ref{eq:def_rhoGJ_1})
only from the Maxwell equation,
$\nabla\cdot\mbox{\boldmath$E$}=4\pi \rho_{\rm r}$, and
the frozen-in condition,
assuming stationarity in the co-moving frame,
namely
$F_{\mu t}+\Omega F_{\mu \varphi}
 = -\partial_\mu \Psi(r,\theta,\varphi-\Omega t) $,
where $F_{\mu\nu}$ represents the field-strength tensor,
$\Psi$ the non-corotational potential,
and $\mu=t,r,\theta,\varphi$
(Hirotani~2006).
That is, equation~(\ref{eq:def_rhoGJ_1})
holds for arbitrary magnetic field, and
is derived irrespective of 
how the current is constructed,
or how the plasmas are collisional or collision-less.

Let us briefly perform a thought experiment.
If the plasma density is large enough,
sufficient collisions allow us to use 
the generalized Ohm's law to describe the current.
In this case, $\rho_{\rm GJ}$ does not diverge 
at the LC,
because the $-(\varpi/\varpi_{\rm LC})^2$ term
in the coefficient of $\rho_{\rm GJ}$ in equation~(\ref{eq:rho_1}) 
comes from the $\mbox{\boldmath$E$}\times\mbox{\boldmath$B$}$ drift,
which is not included in the Ohm's law.
For example, the magnetohydrodynamic approximation,
which uses the Ohm's law to close the equations,
shows that all the physical quantities are well-behaved
at the LC (e.g., Tchekhovskoy et al.~2013).
Next, imagine that the plasmas suddenly escape from the magnetosphere
to become collision-less.
Even in this case, $\rho_{\rm GJ}$ should not be changed at all, 
because $\rho_{\rm GJ}$ is determined only by 
the $\mbox{\boldmath$B$}$ field through equation~(\ref{eq:def_rhoGJ_1}), 
independently from the collisional status of plasmas.
Thus, $\rho_{\rm GJ}$ does not diverge at the LC
also in the collision-less limit.
For example, in the force-free limit,
which adopts the $\mbox{\boldmath$E$}\times\mbox{\boldmath$B$}$ drift
in $\mbox{\boldmath$J$}_\perp$,
no physical quantities diverge or rapidly change
at the LC  (e.g., Spitkovsky 2006),
except for the current sheet, in which the force-free approximation
breaks down.

In general, quantities behave normally across the LC,
without showing any divergence or quick variations.
Thus, the light cylinder gap,
which has an extreme acceleration electric field
(as $10^3$ times stronger than the outer gap),
will not arise in a pulsar magnetosphere, unfortunately.

%
%
%
%

%



\section*{Acknowledgments}
This work is partly supported by 
the Formosa Program between National Science Council  
in Taiwan and Consejo Superior de Investigaciones Cientificas
in Spain administered through grant number 
NSC100-2923-M-007-001-MY3.

%
%
%
%
%
%
\label{lastpage}


\begin{thebibliography}{99}
\bibitem[\protect\citeauthoryear{Aleksic}{2012}(2012)]{alek12} 
  Aleksi\'c, J. {\it et al.} 2012, AA 540, 69
\bibitem[\protect\citeauthoryear{Aliu}{2011}]{aliu11} 
  Aliu, E. Arlen, T., Aune, T., {\it et al.} 2011, Science 334, 69
\bibitem[\protect\citeauthoryear{Atwood et al.}{2009}]{atwood09}
  Atwood, W. B. et al.
  2009, ApJ 697, 1071
\bibitem[\protect\citeauthoryear{Bednarek}{2013}]{bed13} 
  Bednarek, W. 2013, MNRAS, 424, 2079
\bibitem[\protect\citeauthoryear{Cheng et al.}{2000}]{cheng00} 
  Cheng, K. S., Ruderman, M. \& Zhang, L.
  2000, ApJ 537, 964
\bibitem[\protect\citeauthoryear{Goldreich, \&  Julian}{1969}]{gol69} 
  Goldreich, P. \&  Julian, W. H.
  1969, ApJ 157, 869
\bibitem[\protect\citeauthoryear{Hirotani}{2006}]{hir06} 
  Hirotani, K. 2006, MPLA21, 1319
\bibitem[\protect\citeauthoryear{Hirotani}{2011}]{hir11} 
  Hirotani, K. 2011, ApJ, 733, L49
\bibitem[\protect\citeauthoryear{Mestel \& Wang}{1982}]{mest82} 
  Mestel, L., \& Wang, Y. -M. 1982, MNRAS, 198, 405
\bibitem[\protect\citeauthoryear{Spitkovsky}{2006}]{spit06} 
  Spitkovsky, A. 2006, ApJ 648, L51
\bibitem[\protect\citeauthoryear{Tchekkovskoy et al.}{2013}]{tchek13} 
  Tchekhovskoy, A., Spitkovsky, A., Li, J. G. 2013, MNRAS 435, L1
\end{thebibliography}
\end{document}